\documentclass[aps,prc,fleqn,floatfix,twoside,twocolumn,superscriptaddress]{revtex4}

\usepackage[dvipdfm]{graphicx}

\usepackage{amsmath,amssymb,amsfonts}

\usepackage{color}

\begin{document}

\title{Radiative and Collisional  Jet Energy Loss in the Quark-Gluon Plasma at RHIC}

\author{Guang-You Qin}
\author{J\"org Ruppert}

\author{Charles Gale}
\author{Sangyong Jeon}
\author{Guy D. Moore}

\affiliation{Department of Physics, McGill University, Montreal, Quebec, H3A 2T8, Canada}

\author{Munshi G. Mustafa}
\affiliation{Theory Division, Saha Institute of Nuclear Physics, 1/AF Bidhannagar, Kolkata, 700064, India}          

\date{\today}
\begin{abstract}

We calculate and compare bremsstrahlung and collisional energy loss of
hard partons traversing a quark-gluon plasma.  Our treatment of both
processes is complete at leading order in the coupling and accounts for
the probabilistic nature of the jet energy loss. We find that the
nuclear modification factor $R_{AA}$ for neutral $\pi^0$ production in
heavy ion collisions is sensitive to the inclusion of collisional and
radiative energy loss contributions while the averaged energy loss only
slightly increases if collisional energy loss is included for parent
parton energies $E\gg T$. These results are important for the
understanding of jet quenching in Au+Au collisions at $200~{\rm AGeV}$
at RHIC. Comparison with data is performed applying the energy loss
calculation to a relativistic ideal (3+1)-dimensional hydrodynamic
description of the thermalized medium formed at RHIC.

\end{abstract}
\maketitle


{\it Introduction} -- Relativistic heavy ion collisions are designed
to produce and study strongly interacting matter at high temperatures
and densities. Experiments at the Relativistic Heavy Ion Collider (RHIC)
have demonstrated that high $p_T$ hadrons in central $A+A$ collisions
are significantly suppressed in comparison with those in binary $p+p$
collisions, scaled to nucleus-nucleus collisions \cite{Adcox:2001jp,
Adler:2002xw, Gyulassy:1993hr}.
This result has been referred to as jet quenching and has been
attributed to the energy loss of hard $p_T$ partons due to induced gluon
bremsstrahlung in a hot quark-gluon plasma.  
Bremsstrahlung energy loss has been calculated in several theoretical
formalisms before
\cite{Baier:1996kr,Gyulassy:2000er,Kovner:2003zj,Zakharov:2007pj,
Wang:2001if,Arnold:2001ms}.
Recently such bremsstrahlung calculations were implemented in
models employing relativistic ideal (3+1)-dimensional hydrodynamics in
order to calculate the nuclear modification factor $R_{AA}$  of neutral
pions at RHIC \cite{Renk:2006sx,Majumder:2007ae,Qin:2007ys}.
Early estimates of the collisional energy loss which used asymptotic
arguments indicated that the radiative energy loss is much larger than
elastic energy loss \cite{Bjorken:1982tu}.
Zakarhov compared radiative energy loss in the light-cone path integral
approach and collisional energy loss employing the Bjorken method and
concluded collisional energy loss is relatively small in comparison to
the radiative one \cite{Zakharov:2007pj}.  Renk derives phenomenological
limits on radiative {\it vs.} collisional energy loss by considering
quadratic {\it vs.} linear pathlength dependence and concludes that
any elastic energy loss component has to be small \cite{Renk:2007id}.
In contrast, Mustafa and Thoma find that
collisional energy loss has a significant influence on jet
quenching \cite{Mustafa:2003vh,Mustafa:2004dr}. 
Recent studies by Gyulassy and collaborators also 
point in this direction, see e.g. \cite{Adil:2006ei,Wicks:2007mk}.
 
The purpose of this study is to consistently incorporate
collisional and radiative energy loss in the same formalism and to
employ this formalism in a realistic description of energy loss of hard
$p_T$ leading partons in the soft nuclear medium as described by
(3+1)-dimensional  hydrodynamics in $200~{\rm AGeV}$ Au+Au collisions at
RHIC.

\begin{figure}[htb]
\begin{center}
\includegraphics[width=8cm]{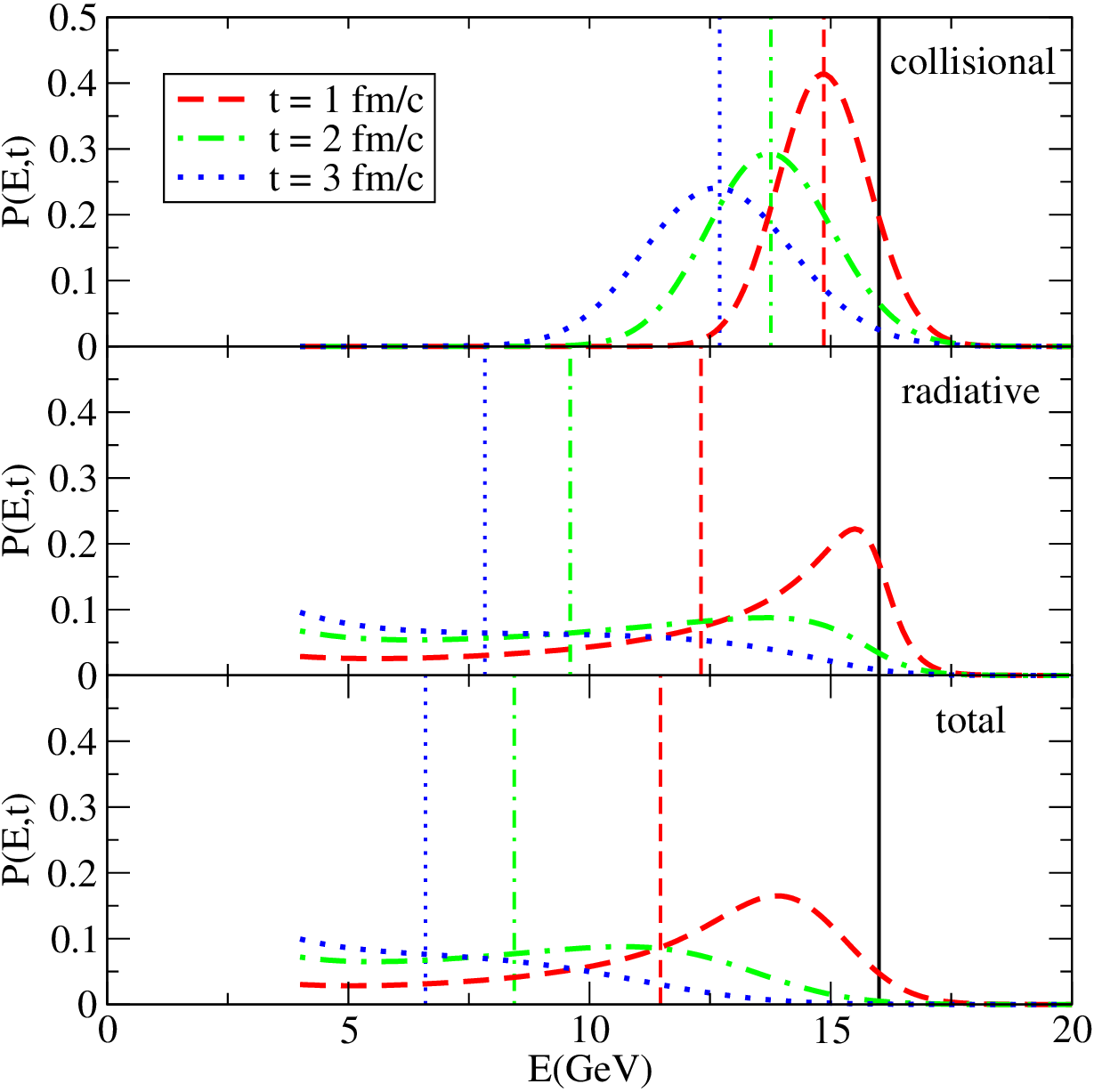}
\end{center}
\caption{(Color online) The evolution of a quark jet with initial energy
  $E_i = 16~{\rm GeV}$ propagating through a static medium of
  temperature $T = 400~{\rm MeV}$, where the vertical lines represent
  the values of mean energy related to the corresponding
  distributions. }
\label{fig1}
\end{figure}

We will emphasize three points  (the first two of which have been
elucidated earlier \cite{Jeon:2003gi} for bremsstrahlung energy loss).
First,  in many previous approaches the {\it average} energy loss is
computed and applied to the primary partons. Bremsstrahlung and
collisional energy loss are not well described by a (path length
dependent) average energy loss alone.  Bremsstrahlung energy loss
is dominated by hard emissions. Therefore, if a sample of partons
initially has the same energy, then after traversing some pathlength of
the medium, the distribution of final energies will be in general broad
and not sharply centered at an average energy. This will be illustrated in
Fig.~\ref{fig1}. While the average energy loss has some value in judging
the importance for observable consequences in jet-quenching, the
evolution of the probability density distributions of partons until
fragmentation is the decisive quantity for such studies. To account for
this we directly evolve the spectrum of partons as they undergo
bremsstrahlung and collisional energy loss.  Second, radiative energy
loss depends on a coherence effect: the Landau-Pomeranchuk-Migdal
(LPM) suppression. While some approaches take the LPM effect as a
parametrically large suppression, this is only true when the parent
parton and the emitted gluon are highly energetic, $E_{\rm parton},
E_{\rm gluon} \gg T$. For small radiated gluon energies $E_{\rm gluon}
\ll E_{\rm parton}$ the LPM suppression can be significantly less, and
those bremsstrahlung events are of significant importance in 
jet quenching \cite{Jeon:2003gi} due to the steeply falling initial
parent parton spectrum.  We therefore employ the Arnold, Moore and
Yaffe (AMY) formalism \cite{Arnold:2001ms} for radiative energy loss to
treat the LPM effect at all energies $E_{\rm gluon} > g_s T$ correctly
up to $O(g_s)$.  Third, while there has been considerable theoretical
effort to improve our understanding of jet modification in the
quark-gluon medium, early jet quenching calculations often relied on an
elementary description of the soft medium. Until recently most jet
quenching calculations used simple medium models only loosely constrained by
the value of bulk observables.
Previously we presented a calculation of $R_{AA}$ in central and
non-central collisions using the AMY formalism and a (3+1) dimensional
hydrodynamical model constrained by soft observables at RHIC
\cite{Qin:2007ys}.  Here we incorporate collisional energy loss into this
framework.

{\it Brief review of the formalism} --
For details of the calculation of the initial distributions of hard
partons in the early stage of the collision and the subsequent
propagation through the hot and dense quark gluon plasma as well as the
fragmentation we refer the reader to \cite{Qin:2007ys} and references
therein.  We concentrate here on the incorporation of collisional energy
loss in the formalism.

The jet momentum distribution $P(E,t)=\frac{dN(E,t)}{dE}$ evolves
in the medium according to a set of coupled 
Fokker-Planck type rate equations, which have the following generic form
 \cite{Qin:2007ys}:
\begin{equation}
\label{jet-evolution-eq}
\frac{dP(E)}{dt}= \int_{-\infty}^{\infty} \!\!\!\!\!\!d\omega
\!\left[P(E{+}\omega) \frac{d\Gamma(E{+}\omega,\omega)}{d\omega} 
 - P(E)\frac{d\Gamma(E,\omega)}{d\omega}\!\right]
\end{equation}
where ${d\Gamma(E,\omega)}/{d\omega}$ is the transition rate for
processes where partons of energy $E$ lose energy $\omega$.  The
$\omega<0$ part of the integration incorporates processes which increase
a particle's energy.  The radiative part of the transition rate is
discussed in \cite{Jeon:2003gi, Turbide:2005fk,Qin:2007ys}.

Now we must add the contribution from collisional energy loss.  Compared
to radiative loss, collisional losses are more dominated by small energy
transfers; the contribution to the
mean energy loss rate $dE/dt$ from elastic collisions is
logarithmically sensitive to large energy transfers, while the
radiative contribution is power-law dominated by large radiations.
Therefore it should be an adequate procedure to approximate elastic
collisions by a mean energy loss, {\it provided} we include a
momentum diffusion term as dictated by detailed balance.

The leading order mean collisional energy loss rate is
\begin{eqnarray}
\label{dEdt_omega}
\frac{dE}{dt} &=& \frac{g_k}{2E} \int \frac{d^3k}{(2\pi)^3 2k} \int
\frac{d^3p'}{(2\pi)^3 2E'} \int \frac{d^3k'}{(2\pi)^3 2k'}
\nonumber\\ && \times (2\pi)^4\delta^4(P+K-P'-K') \nonumber\\ 
&&  \times (E-E') |\bar{M}|^2  f(k) [1\pm f(k')] \,,
\end{eqnarray}  
where $f$ is the thermal distribution of the medium
partons. $|\bar{M}|^2$ is the $t$-channel scattering matrix element
squared calculated in leading  order, and $g_k$ is the degeneracy factor for
the initial thermal partons. Eq.~(\ref{dEdt_omega}) is infrared
logarithmic divergent, screened by plasma effects which are incorporated
by including hard thermal loop corrections for soft momenta $\sim gT$. 
The resulting differential energy loss $dE/dt|_{ab}$
for the scattering of a light hard parton $a$ off a
soft parton $b$ are \footnote{The collisional energy loss as
  calculated by  \cite{Thomas:1991ea} differs by constant terms in the
  brackets in Eqs.~(\ref{dEdt}) since $u \approx -s$ was employed in the
  numerator of the matrix elements squared for the hard scattering
  there. The differences are not of phenomenological importance.}:
\begin{eqnarray}
\label{dEdt}
\left.\frac{dE}{dt}\right|_{qq} &=& \frac29 n_f \pi \alpha_s^2 T^2 
 \left[\ln \frac{ET}{m_g^2} +
c_f + \frac{23}{12} + c_s \right], \ \ \ \ \ \  \nonumber \\
\left.\frac{dE}{dt}\right|_{qg} &=& \frac43 \pi \alpha_s^2 T^2 
 \left[\ln \frac{ET}{m_g^2} +
c_b + \frac{13}{6} + c_s \right],  \ \ \ \ \ \  \nonumber \\
\left. \frac{dE}{dt}\right|_{gq} &=& \frac12 n_f \pi \alpha_s^2 T^2 
\left[\ln \frac{ET}{m_g^2} +
c_f + \frac{13}{6} + c_s \right],  \ \ \ \ \ \  \nonumber \\
\left.\frac{dE}{dt}\right|_{gg} &=& 3 \pi \alpha_s^2 T^2 
\left[\ln \frac{ET}{m_g^2} +
c_b + \frac{131}{48} + c_s \right], \ \ \ \ \ \
\end{eqnarray}
where $c_b = - \gamma_E + {\zeta'(2)/ \zeta(2)}$, $c_f=c_b+\ln(2)$, and
$c_s \approx -1.66246$ are constants and
$m_g^2 = 2 \pi \alpha_s T^2 (1 + n_f/6)$ is the thermal gluon
mass \cite{GaleKapusta}.

We can incorporate these $dE/dt$ results in Eq.~(\ref{jet-evolution-eq})
by introducing the drag term, $(dE/dt)dP(E)/dE$, and
the diffusion term, $T(dE/dt)d^2P(E)/dE^2$.  We discretize
Eq.~(\ref{jet-evolution-eq}), such that $\int d\omega \rightarrow \Delta\omega
\sum_{\omega = n\Delta \omega}$, and 
\begin{eqnarray}
\Gamma(E+\Delta\omega,\Delta\omega) &=& (1+f_B(\Delta\omega))(\Delta
\omega)^{-1} dE/dt \, , \nonumber \\
\Gamma(E,-\Delta \omega) & = & f_B(\Delta\omega) (\Delta\omega)^{-1}
dE/dt \,,
\end{eqnarray}
which yields the right energy loss rate and preserves detailed balance.

{\it Results} -- In order to illustrate how collisional and radiative
energy loss influence the time evolution of the leading parton
distributions, we first consider a static infinite medium with
$T=400~{\rm MeV}$ and $\alpha_s=1/3$ and an initial single light quark
jet of energy $E_i=16~{\rm GeV}$ propagating through it.

In Fig. \ref{fig1} we compare the evolution of the jet parton
distribution $P(E,t)$ under three different approximations: (1) with
only collisional energy loss, (2) with only radiative energy loss
(already calculated in \cite{Jeon:2003gi}), and (3)  with both
energy loss mechanisms. The first moment in energy of
these distributions defines the mean energy (indicated by vertical
lines) and indicates the average energy loss.  The figure indicates as
expected that radiation lead to a larger mean energy loss
than with elastic collisions only.
As pointed out earlier, small differences in the average energy loss do
not necessarily  imply small differences in the parton distributions. While the time evolution of $P(E,t)$ in case (3) resembles
qualitatively the case (2) in which only radiative energy loss has been
considered, quantitative differences especially at energies closer to
$E_i$ can be large.

We will model the behavior of the quark-gluon medium using relativistic
fluid dynamics, which has been shown to give a good description of bulk
properties at RHIC.  We use a fully (3+1) dimensional
hydrodynamical model for the description of $200~{\rm AGeV}$ Au+Au
collisions at RHIC \cite{Nonaka:2006yn}.  The initial momentum
distribution of jets ${dN^j}/{d^2p^j_T dy}|_{\rm ini}$ is computed from
pQCD in the factorization formalism, for details see
\cite{Lai:1999wy,Eskola:1998df,Qin:2007ys}.  The final hadron spectrum
${dN^h}/{d^2p_Tdy}$  at high $p_T$ is obtained by the fragmentation of
jets in the vacuum after their passing through the (3+1) dimensional
expanding medium
\begin{eqnarray}
\label{hadron_distr} 
\hspace{-0.15in} \frac{dN^h}{d^2p_Tdy} &=& \sum_{j} \int d^2\vec{r}_\bot
\mathcal{P}(b, \vec{r}_\bot) \int \frac{ dz_j}{z_j^2} {D_{h/j}(z_j,Q_F)}
\nonumber \\ && \times
\left. \frac{dN^{j}(b,\vec{r}_\bot)}{d^2p^j_Tdy}\right|_{\rm fin},
\end{eqnarray}
where ${dN^{j}(b,\vec{r}_\bot)}/{d^2p^j_Tdy}|_{\rm fin}$ is the final
momentum distribution of the jet initially created at transverse
position $\vec{r}_\bot$ after passing through the medium. This
distribution is calculated for every specific path through the medium
by solving Eq.(\ref{jet-evolution-eq})
incorporating
collisional and radiative energy loss. The fragmentation function
${D_{h/j}(z_j,Q_F)}$ \cite{Kniehl:2000fe} gives the average multiplicity
of the hadron $h$ with momentum fraction $z_j=p_T/p^j_T$ produced
from a jet $j$ at fragmentation scale $Q_F$.
$\mathcal{P}(b,\vec{r}_\bot)$ is the initial jet density distribution at
the transverse position $\vec{r}_\bot$ in collisions with  impact
parameter $\vec{b}$.  For further details see \cite{Qin:2007ys} where
radiative energy loss has been studied in an analogous manner.

The final hadron spectrum directly enters the calculation of the nuclear
modification factor $R_{AA}$ which is defined as the ratio of the hadron
yield in A+A collisions to that in p+p interactions scaled by the number
of binary collisions $N_{\rm coll}$:
\begin{eqnarray}
R^h_{ AA}(b,\vec{p}_T,y) &=& \frac{1}{N_{\rm coll}(b)}
\frac{{dN^h}(b)/{d^2p_Tdy}} {{dN^h_{ pp}}/{d^2p_Tdy}}.
\end{eqnarray}

Once temperature evolution is fixed by the initial conditions and
subsequent evolution of the (3+1) dimensional expansion, the strong
coupling constant $\alpha_s$ is the only quantity which is not uniquely
determined by the model. In this work we take its value to be constant
 at $\alpha_{\rm s}=0.27$, which reproduces the most central data 
\footnote{Recently  there has been some discussion about effects of
  running coupling on collisional energy loss \cite{Peshier}, especially
  at high energies $E \rightarrow \infty$.   However, in such calculations the results are found to be sensitive to the choice of input parameters  \cite{Wicks:2007mk}. }.
 
\begin{figure}[htb]
\begin{center}
\includegraphics[width=8cm]{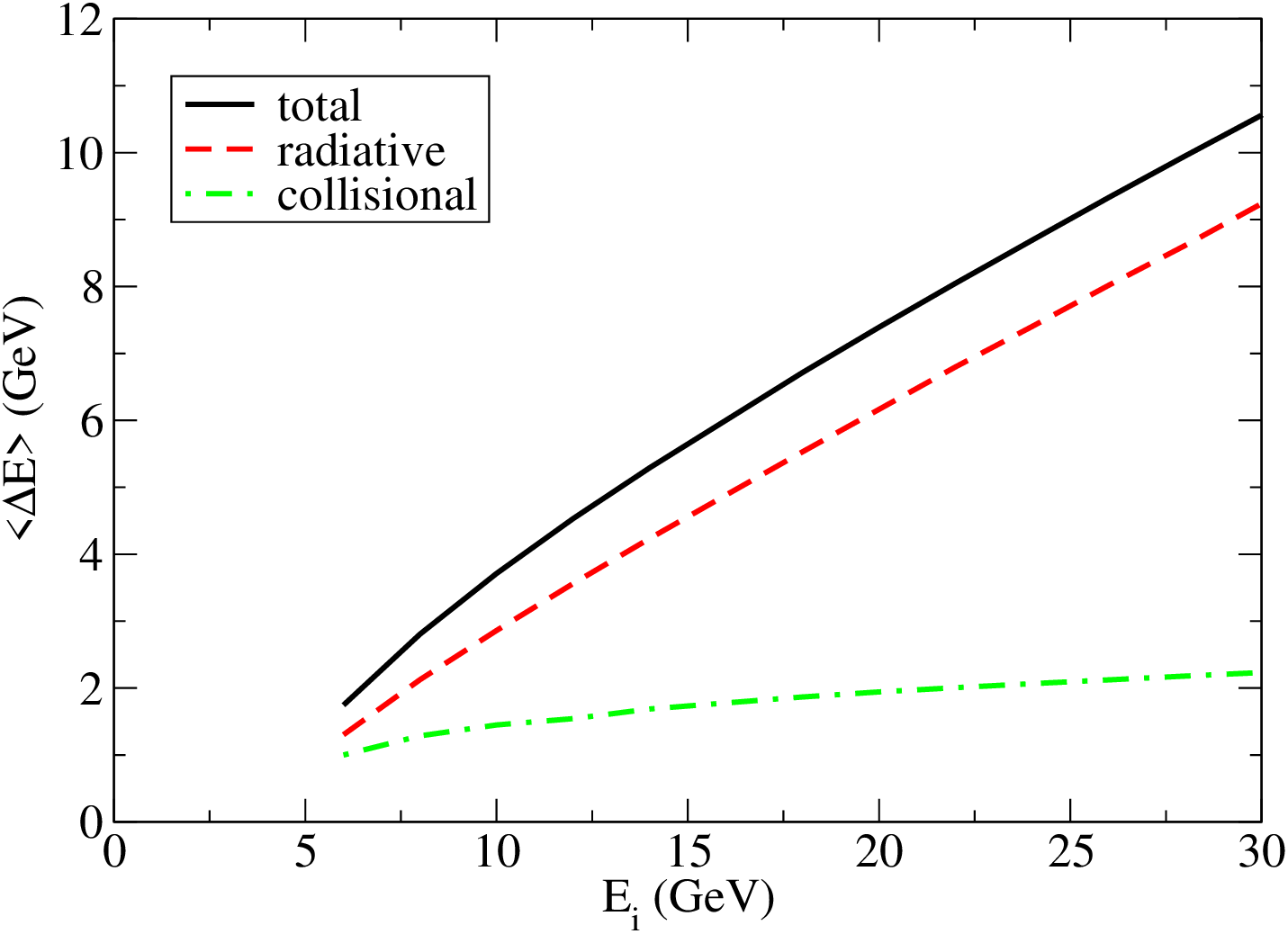}
\end{center}
\caption{(Color online) The mean energy loss of a quark jet with initial
  energy $E_i$ passing through the nuclear medium created in central
  collisions ($b = 2.4~{\rm fm}$) at RHIC. The jet starts from the
  center of the medium and propagates in plane.}
\label{fig2}
\end{figure}

\begin{figure}[htb]
\begin{center}
\includegraphics[width=8cm]{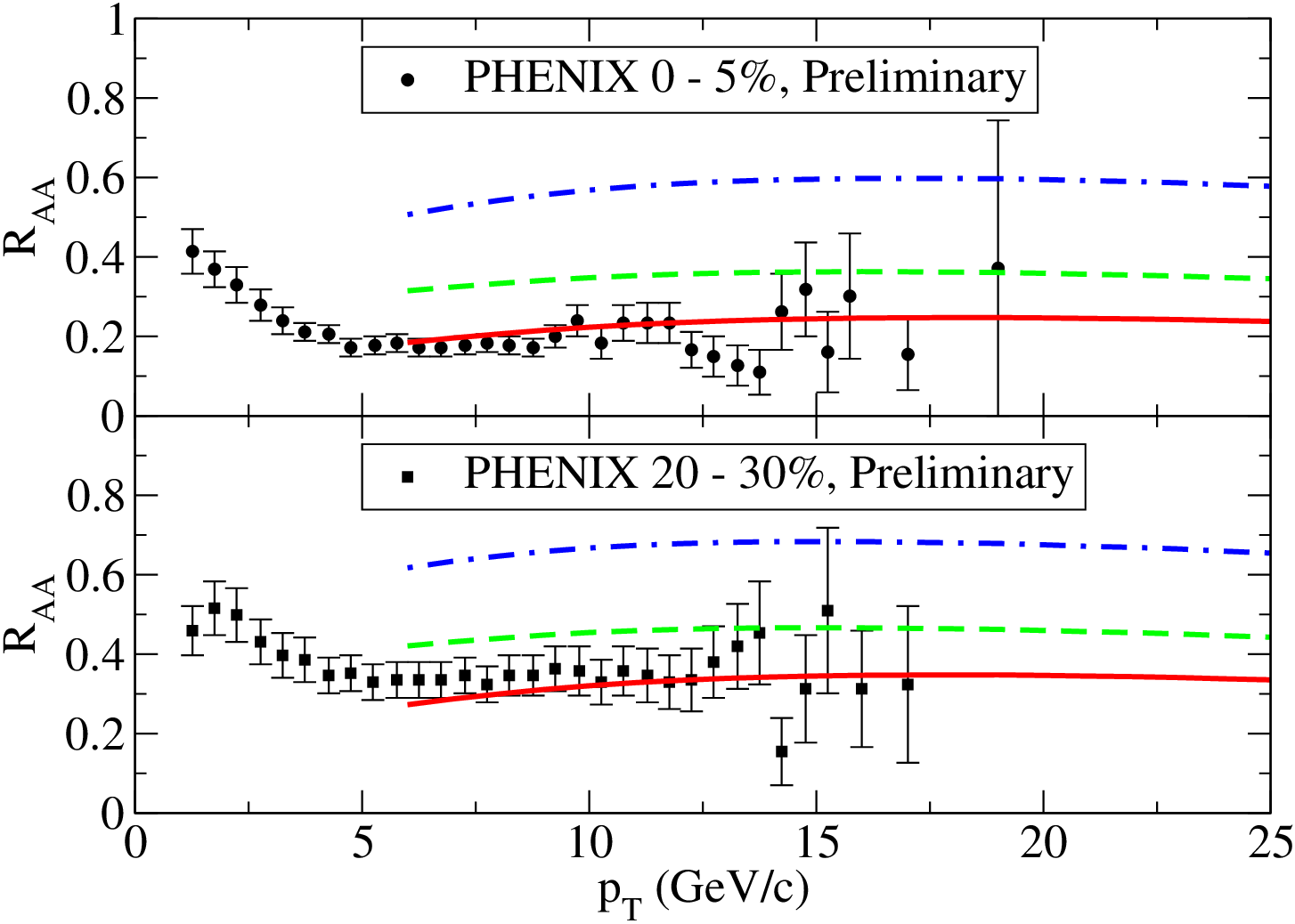}
\end{center}
\caption{(Color online) Nuclear suppression factor $R_{AA}$ for neutral
  pions in central and mid-central collisions. 
  Here
  the dashed curves includes only radiative energy loss, 
  the dash-dotted curve includes only collisional energy loss 
  and the solid curve includes both
  radiative and collisional energy loss.}
\label{fig3} 
\end{figure}

In Fig.\ \ref{fig2} we show the mean energy loss of quark jets passing through the nuclear
medium created in central collisions ($b=2.4$ fm) at RHIC, 
as a
function of their initial energy $E_i$ . In this figure, the jets are
assumed to be created at the center and propagating along the in-plane
direction.  In agreement with \cite{Zakharov:2007pj} we find that the
average energy loss is not strongly changed by accounting for elastic
collisions.  In Fig.\ \ref{fig3} we present the calculation of $R_{AA}$
for neutral pions measured at mid-rapidity for two different impact
parameters, 2.4 fm and 7.5 fm, compared with PHENIX data for the most
central  (0-5\%) and mid-central (20-30\%) collisions
\cite{Adler:2002xw}.  We present $R_{AA}$ for purely collisional (1) and
purely radiative (2) energy loss, as well as the combined case (3).  One
finds that while the shape of $R_{AA}$ for case (3) is not strongly
different from case (2) that has only radiative energy loss, the
overall magnitude of $R_{AA}$ changes significantly.  
We checked
(comparison not shown) that the stronger influence on $R_{AA}$ stems
from the differences in the evolution of the parton distributions in
case (2) and (3). This has already been discussed in the static case
(compare Fig.\ \ref{fig1}). The magnitude of $R_{AA}$ is therefore sensitive
to the actual form of the parton distribution functions at fragmentation
and not only to the average energy loss of single partons
(compare Fig.\ \ref{fig2}).  
In \cite{Qin:2007ys}, the observational
implications on $R_{AA}$ measurements due to only radiative energy loss
were studied.  Recalculating $R_{AA}$ versus reaction plane including
elastic collisions  we found only small differences (after adjusting the coupling strength from $\alpha_s=0.33$ to
$\alpha_s=0.27$) in the shape of $R_{AA}$ as a function of $p_T$ and the
azimuth.
 
{\it Conclusions} -- We calculated collisional and radiative jet energy
loss of hard partons in the hot and dense medium created at RHIC. We
treated the LPM effect using the AMY formalism
\cite{Arnold:2001ms} and treated collisions using a drag plus diffusion
term reproducing $dE/dt$ and detailed balance.
While we find in accordance with \cite{Zakharov:2007pj} that the
additionally induced average energy loss due to elastic collisions
is small in comparison to the radiative one, the time
evolution of the parton distributions $P(E,t)$ in both cases differ
significantly. This  is especially true for energies close to the initial
parton energy. Since the initial parton spectrum is steeply falling,
stronger differences in $R_{AA}$ can result. We find that the inclusion
of collisional energy loss significantly influences the quenching power
quantified as the overall magnitude of neutral pion  $R_{AA}$ at RHIC,
but that the shape as a function of $p_T$ 
does not show a strong sensitivity.
We emphasize that the description of $R_{AA}$ is not enough to prove the
consistency of a specific energy loss mechanism with data, if
assumptions about the medium evolution can be rather freely
adjusted. Therefore folding the jet energy loss mechanism with a
dynamical evolution model which has been well tested to describe soft
observables is necessary.

{\it Acknowledgments} --
We thank C. Nonaka and S. Bass for providing their hydrodynamical
evolution calculation \cite{Nonaka:2006yn}. This work was supported in part by
the Natural Sciences and Engineering Research Council of Canada, by the
McGill-India Strategic Research Initiative, and by the Fonds Nature et Technologies of Quebec.


\end{document}